\documentclass[10pt,dvips]{article}
\usepackage{amssymb,amsfonts,amsmath,latexsym}
\usepackage{graphics,graphicx,epsf}
\newcommand{\be}{\begin{equation}}
\newcommand{\ee}{\end{equation}}
\newcommand{\bea}{\begin{eqnarray}}
\newcommand{\eea}{\end{eqnarray}}
\newcommand{\ba}{\begin{array}}
\newcommand{\ea}{\end{array}}
\newcommand{\bt}{\begin{tabular}}
\newcommand{\et}{\end{tabular}}

\newcommand{\fr}{\frac}
\newcommand{\ci}{\cite}
\newcommand{\cl}{\centerline}
\newcommand{\bs}{\bigskip}

\newcommand{\vs}{\vspace}

\newcommand{\en}{\eqno}

\newcommand{\fns}{\footnotesize}

\newcommand{\bbib}{\begin{thebibliography}}
\newcommand{\ebib}{\end{thebibliography}}

\newcommand{\und}{\underline}

\begin{document}
\bs
\cl{\bf EFFECTIVE CONDUCTIVITY OF 2D
ISOTROPIC} \cl{\bf TWO-PHASE SYSTEMS IN MAGNETIC FIELD}

\bs

\cl{\bf S.A.Bulgadaev \footnote{e-mail: bulgad@itp.ac.ru}, 
\bf F.V.Kusmartsev \footnote{e-mail: F.Kusmartsev@lboro.ac.uk}}

\bs \cl{\fns Landau Institute for Theoretical Physics,
Chernogolovka, Moscow Region, Russia, 142432} \cl{\fns
Department of Physics, Loughborough University, Loughborough, LE11
3TU, UK}

\bs

\begin{quote}
\footnotesize{ Using the linear
fractional transformation, connecting effective conductivities $\hat \sigma_{e}$
of isotropic two-phase systems with and without magnetic field,
explicit approximate expressions for $\hat \sigma_{e}$ in a magnetic field are obtained. They allow to describe $\hat \sigma_{e}$ of various inhomogeneous media at arbitrary phase concentration x and magnetic fields. 
The x-dependence plots of $\hat \sigma_{e}$ at some values of inhomogeneity and magnetic field are constructed. Their behaviour is qualitatively compatible with the
existing experimental data.
The obtained results are applicable
for different two-phase systems (regular and nonregular as well as
random), satisfying the symmetry and self-duality conditions,
and admit a direct experimental checking.
}
\end{quote}

\bs
\cl{PACS: 75.70.Ak, 72.80.Ng, 72.80.Tm, 73.61.-r}
\bs

\underline{1. Introduction}

\bs

Last time, under investigation of magneto-resistive properties of
 new materials, which are connected with the high-temperature superconductivity
(such as oxide materials with the perovskite type  structure),
it was established that they often have unusual transport properties. For example,
the magnetoresistance becomes very large (the so called colossal magnetoresistance
in such materials as manganites) \ci{1} or grows approximately linearly with magnetic field up to very high fields (in silver
chalcogenides) \ci{2}. There is an opinion that these properties take place due to
phase inhomogeneities  of these materials \ci{2}.
For this reason a calculation of the effective conductivity $\sigma_{e}$ of inhomogeneous heterophase systems without and with magnetic field at arbitrary partial conductivities and phase concentrations is very important problem. Unfortunately, the existing effective medium
approximations (EMA) cannot give an explicit simple formulas for magneto-resistivity
convenient for description of the experimental results
in a wide range of partial parameters even for two-phase random systems \ci{3}.
Such formulas can be obtained for some systems only in high magnetic field \ci{3,4}.

The situation is much better for 2D inhomogeneous systems in the perpendicular magnetic field. Here, due to their exact duality properties, not only a few exact results have been obtained for the effective conductivity  of random inhomogeneous systems [5-10], but also the transformations, connecting effective conductivities of isotropic two-phase inhomogeneous systems with and without magnetic field, have been constructed \ci{9,11}. This transformation, in principle, permits to obtain  the explicit expressions for $\sigma_e$ in a magnetic field, if the corresponding expressions are known for $\sigma_e$ without magnetic field. Such explicit approximate expressions for $\sigma_e$ have been obtained earlier in 
some limiting cases, for example, at small concentration of one phase, for weakly inhomogeneous media \ci{11}, for inclusions of super- or non-conducting phases
[6-8,11]. In this letter we will present the explicit approximate expressions for $\hat \sigma_e$ at arbitrary phase concentrations and in a wide region of partial conductivities, using the full dual transformation recently constructed in \ci{10} and the corresponding expressions for $\sigma_e$ at ${\bf H}=0$ from \ci{12}.
We will give also the x-dependence plots of $\hat \sigma_e$ at some characteristic values of magnetic field ${\bf H},$ which show very interesting behaviour, compatible with experimental results from \ci{1,2}.

\bs
\und{2. Dual transformation between systems with ${\bf H}\ne 0$ and ${\bf H}= 0$ }
\bs

The effective conductivity of two-phase isotropic systems in a
magnetic field has the following form
$$
\hat \sigma = \sigma_{ik} = \sigma_d \delta_{ik} + \sigma_t
\epsilon_{ik}, \quad \sigma_d ({\bf H}) = \sigma_d (-{\bf H}),
\quad \sigma_t ({\bf H}) = -\sigma_t (-{\bf H}), 
\en(1)
$$
here $\delta_{ik}$ is the Kronecker symbol, $\epsilon_{ik}$ is the unit antisymmetric tensor.
The effective conductivity $\hat \sigma_{e}$ of $2$-phase self-dual (random
or regular symmetric) systems with the partial conductivities
$\sigma_{id}, \; \sigma_{it} \;(i = 1,2)$ (we assume that
$\sigma_{id} \ge 0$) and concentrations $x_i$  must be a symmetric
function of pairs of arguments ($\hat \sigma_i, x_i$) and a
homogeneous (a degree 1) function of $\sigma_{di,ti}.$ For this
reason it is invariant under permutation of pairs of partial
parameters
$$
\hat \sigma_{e}(\hat \sigma_1, x_1|\hat \sigma_2, x_2) = \hat
\sigma_{e}(\hat \sigma_2, x_2|\hat \sigma_1, x_1).
\en(2)
$$
The effective conductivity of $2$-phase systems must also
reduce to some partial $\hat \sigma_i$, if $x_i = 1 \;(i=1,2).$

Futher it will be more
convenient to use the complex representation for coordinates and
vector fields \ci{13}
$$
z=x+iy, \quad j= j_x+ij_y, \quad e=e_x+ie_y,  \quad \sigma =
\sigma_d + i\sigma_t.
$$
The complex conductivity transforms as \ci{5}
$$
\sigma' = T(\sigma) = \fr{c \sigma -ib}{-id \sigma + a}, 
\en(3)
$$
where $a,b,c,d$ are the real numbers. This transformation generalizes the inversion transformations of systems without
magnetic field. Thus, the dual transformations (DT)
in systems with magnetic field have a more richer structure due to
the fact that they are connected with some subgroup of group of
linear fractional transformations, conserving the imaginary axis in the complex conductivity plane [5,7,9].
The transformation $T$ depends on
3 real parameters (since one of 4 parameters can be factored
due to the fractional structure of T). There are various ways to
choose 3 parameters. In our treatment it will be convenient to
factor $d.$ This gives 3 parameters $\bar a = a/d, \bar b=
b/d,\bar c= c/d,$ determining a transformation $T.$ 

The transformation (3) has the following form in terms of
conductivity components $\sigma_d$ and $\sigma_t$
$$
\sigma_d' = \sigma_d \fr{ac + bd}{(d \sigma_d)^2 + (a+ d
\sigma_t)^2} = \bar c \sigma_d \fr{\bar a  + {\bar b}/{\bar c}}{(
\sigma_d)^2 + (\bar a + \sigma_t)^2},
$$
$$ \sigma_t' = \fr{cd \sigma_d^2 + (a+d
\sigma_t)(c\sigma_t -b)}{(d \sigma_d)^2 + (a+ d \sigma_t)^2} =
{\bar c} \fr{ \sigma_d^2 + ({\bar a} + \sigma_t)(\sigma_t -b/c)}{(
\sigma_d)^2 + ({\bar a}+ \sigma_t)^2}. 
\en(4)
$$
These DT allow to construct the transformation, connecting effective 
conductivities of two-phase systems with magnetic field and without it. 
Analogous connections have been found firstly on a basis of solutions
of the corresponding Laplace and boundary equations \ci{11}, but it has some ambiguity problems with a determination of the parameters of the intermediate artificial system. 
Later, this transformation has been constructed directly from the DT under two simplifying conjectures  in \ci{9}. Recently, it was  constructed in the full form, using all 3 parameters and explicitly reproducing the known exact
values for $\sigma_e,$ in \ci{10}. The parameters of such transformation (let
us call it $T_h$, we will also omit the bars over its parameters) $a,b'=b/c,c$ depend on the partial conductivities and have the following form
$$
a_{\pm} = \fr{|\sigma_2|^2 - |\sigma_1|^2 \pm
\sqrt{B}}{2(\sigma_{1t} - \sigma_{2t})}, \quad b'_{\pm} =
\fr{|\sigma_1|^2 - |\sigma_2|^2 \pm \sqrt{B}}{2(\sigma_{1t} -
\sigma_{2t})},\quad c=-a,
$$
$$
B = [(\sigma_{1t} - \sigma_{2t})^2 + (\sigma_{1d} -
\sigma_{2d})^2] [(\sigma_{1t} - \sigma_{2t})^2 + (\sigma_{1d} +
\sigma_{2d})^2], 
\en(5)
$$
where $|\sigma_i|^2 = \sigma_{id}^2 + \sigma_{it}^2,$ and, evidently, $B\ge 0.$ 
The diagonal (or real) parts of $\sigma_i$  transform under $T_h$
as
$$
\sigma_{id}' =  \sigma_{id} \fr{c(a + b')}{(\sigma_{id})^2 + (a+
\sigma_{it})^2} =   \fr{c \sigma_{id}}{\sigma_{ai}}, \quad
\sigma_{ai} = a + \sigma_{it}. \en(6)
$$
The parameters $a,b,c$ satisfy also the additional relations \ci{10}
$$
A = \left[1+\left(\fr{\sigma_{1t} - \sigma_{2t}}{\sigma_{1d} +
\sigma_{2d}}\right)^2\right]^{1/2} = \fr{(a+b')(\sigma_{a1}
\sigma_{a2})^{1/2}}{\sigma_{1d} \sigma_{2d} + \sigma_{a1}
\sigma_{a2}}, \en(7)
$$
$$
\fr{\sigma_{1d} \sigma_{2t} + \sigma_{2d} \sigma_{1t}}{\sigma_{1d}
+ \sigma_{2d}}= c\fr{{\sigma'_e}^2 -a b'}{{\sigma'_e}^2 + a^2} =
c \fr{\sigma_{1d} \sigma_{2d} - (b'/a) \sigma_{a1}
\sigma_{a2}}{\sigma_{1d} \sigma_{2d} + \sigma_{a1}
\sigma_{a2}}, \en(8)
$$
which ensure the reproduction of the exact results for $\sigma_e$ at the equal
phase concentrations $x_1=x_2=1/2.$
The relations (7) and (8) give us a highly nontrivial check of a
self-consistency of the transformation. The direct check
of them is a rather complicated task. Fortunately, as it was conjectured  in \ci{9} and shown in \ci{10}, $T_h$ transforms a circumference (see fig.1.) 
\begin{figure}[t]
\cl{\input epsf \epsfxsize=8cm \epsfbox{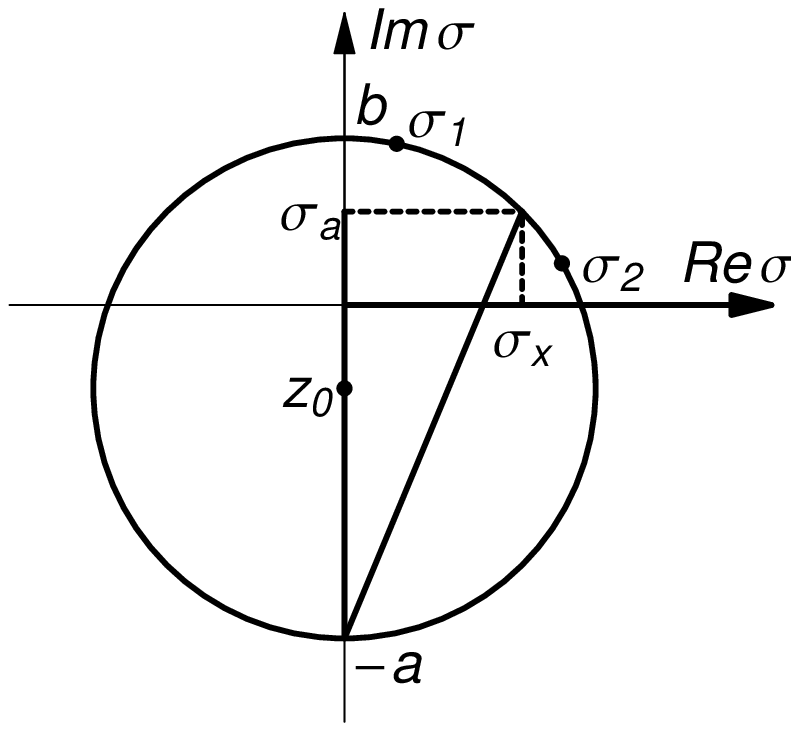}}

\vs{0.3cm}

{\small  Fig.1.  A schematic picture of the chord, defining a
geometrical sense  of the factor $A$ from the expression for the
real part of the exact $\sigma_e$ .}
\end{figure}
with a radius $R,$ centered on the imaginary axis at $iz_0$ and passing through  the two points $\sigma_1$ and $\sigma_2,$ where
$$
R = \fr{|a+b'|}{2} = \fr{\sqrt{B}}{2|\sigma_{1t} - \sigma_{2t}|}, 
\quad z_0 = \fr{-a+b'}{2} = \fr{|\sigma_1|^2 - |\sigma_2|^2}{2(\sigma_{1t} -
\sigma_{2t})},
\en(9)
$$
into the real axis and the real axis into this circumference.
All equalities necessary for fulfillment of (7),(8) correspond to the
known equalities between lengths of various chords, their
projections on a diameter and the radius of the circles. 
For example, the equality (7) takes the form
$$
A= \fr{2R\sigma_a }{\sigma_a^2 + \sigma_{1d} \sigma_{2d}} =
1/(1-\delta/2R\sigma_a), \quad \sigma_a^2 = \sigma_{a1}
\sigma_{a2},
\en(10)
$$
where $\sigma_a^2$ is a squared geometrical average of
$\sigma_{ai}, \; (i=1,2)$ (this interpretation is possible due to the fact that $\sigma_{ai}$ always have the same sign \ci{9}), and $\delta$ defines a  difference
between $\sigma_x^2 = 2R\sigma_a - \sigma^2_a,$ a squared real
projection of the chord, having an imaginary projection
$\sigma_a,$  and a squared geometric average of $\sigma_{id}$
$$
\delta = \sigma_x^2 - \sigma_{1d} \sigma_{2d}.
\en(11)
$$

\und{3. Inhomogeneous systems with compact inclusions}

\bs

Now, having all formulas for the transformation, one can construct
the explicit expressions for $\sigma_e$ of inhomogeneous system in
a magnetic field. Below, in order to simplify a view of the subsequent formulas, we will omit the subindex $d$ in the partial diagonal parts $\sigma_{id} \; (i=1,2).$ Firstly we consider the case of compact
inclusions of one phase into another. The corresponding effective
conductivity $\sigma_e$ has the next form ($x_1=x, x_2 = 1-x$) \ci{12}
$$
\sigma_e(\{\sigma\}, \{x\}) = \sigma_1^{x} \sigma_2^{1-x} 
\en(12)
$$
Substituting  the primed partial conductivities into (12), one obtains the diagonal
primed effective conductivity 
$$ 
\sigma'_{ed} (\{\sigma\}, \{x\}) = c \left(\fr{\sigma_1}{\sigma_{a1}}\right)^{x} 
\left(\fr{\sigma_2}{\sigma_{a2}}\right)^{1-x}
\en(13)
$$
Then, substituting $\sigma'_{ed}$ into (4) and remembering that the primed transverse effective conductivity $\sigma'_{et}=0$, one obtains for the diagonal part $\sigma_{ed}$ of the effective conductivity of these systems in
a magnetic field the following expression 
$$
\sigma_{ed} (\{\sigma\}, \{x\}) =
\fr{\sigma'_{ed}(ac+b)}{(\sigma'_{ed})^2 + a^2} =
 \fr{(a+b')\left(\fr{\sigma_{1}}{\sigma_{a1}}\right)^{x_1}
\left(\fr{\sigma_{2}}{\sigma_{a2}}\right)^{x_2}}{1 +
\left(\fr{\sigma_{1}}{\sigma_{a1}}\right)^{2x_1}
\left(\fr{\sigma_{2}}{\sigma_{a2}}\right)^{2x_2}}. 
\en(14)
$$
One can check, using the relation (7), that (14) correctly
reduces to $\sigma_i,$ when $x_i =1$ and to the exact formula for
$\sigma_e$ at $x_1=x_2=1/2.$ For example, at $x_1 = 1, x_2 = 0$ (14)
reduces to
$$
\sigma_{ed} (\{\sigma\}, x_1=1)  = \fr{\sigma_{1}}{\sigma_{a1}}  \fr{(a + b')}{1 +
\left(\fr{\sigma_{1}}{\sigma_{a1}}\right)^{2}} = \sigma_1,
$$
where at the last step we have used the relation (7) at $\sigma_1
=\sigma_2.$

Analogously, for the transverse part $\sigma_{et}$ one obtains
$$
\sigma_{et} (\{\sigma\}, \{x\}) = c \fr{(\sigma'_{ed})^2 -a
b'}{(\sigma'_{ed})^2 + a^2}  =  \fr{b' - a \left(\fr{\sigma_{1}}{\sigma_{a1}}\right)^{2x_1}
\left(\fr{\sigma_{2}}{\sigma_{a2}}\right)^{2x_2}}{1+ \left(\fr{\sigma_{1}}{\sigma_{a1}}\right)^{2x_1} \left(\fr{\sigma_{2}}{\sigma_{a2}}\right)^{2x_2}}. 
\en(15)
$$
Again, using now the relation (8), one can check that (15) correctly
reproduces boundary values at $x_i=1\; (i=1,2)$ as well as the exact value at equal
concentrations $x_1=x_2=1/2.$

\bs
\und{4. Inhomogeneous systems with a "random parquet" structure}

\bs

In this section we present analogous formulas for 2D isotropic
inhomogeneous systems with the structure of the inhomogeneities of
the "random parquet" type \ci{12}. The effective conductivity of such
systems without a magnetic field is \ci{12}
$$
\sigma_{e} (\{\sigma\}, \{x\}) = \sqrt{\langle \sigma \rangle
/\langle \sigma^{-1} \rangle} = \sqrt{\sigma_1 \sigma_2}
\left(\fr{x_1 \sigma_1 + x_2 \sigma_2}{x_1 \sigma_2 + x_2 \sigma_1}\right)^{1/2}
\en(16)
$$
Then the diagonal part of the primed effective
conductivity $\sigma'_e$ has the form
$$
\sigma'_{ed} (\{\sigma\}, \{x\}) = c \left(\fr{\sigma_{1}}{\sigma_{a1}}
\fr{\sigma_{2}}{\sigma_{a2}}\right)^{1/2} 
\left(\fr{x_1 \left(\fr{\sigma_{1}}{\sigma_{a1}}\right) + x_2
\left(\fr{\sigma_{2}}{\sigma_{a2}}\right)}{x_1
\left(\fr{\sigma_{2}}{\sigma_{a2}}\right) + x_2
\left(\fr{\sigma_{1}}{\sigma_{a1}}\right)}\right)^{1/2} \en(17)
$$
Its substitution into (4) gives for the diagonal part of the effective
conductivity $\sigma_{ed}$
$$
\sigma_{ed} (\{\sigma\}, \{x\}) = (a+b')
\fr{\left(\fr{\sigma_{1}}{\sigma_{a1}}
\fr{\sigma_{2}}{\sigma_{a2}}\right)^{1/2} \left(\fr{x_1 \left(\fr{\sigma_{1}}{\sigma_{a1}}\right) + x_2
\left(\fr{\sigma_{2}}{\sigma_{a2}}\right)}{x_1
\left(\fr{\sigma_{2}}{\sigma_{a2}}\right) + x_2
\left(\fr{\sigma_{1}}{\sigma_{a1}}\right)}\right)^{1/2}}{1+ \fr{\sigma_{1}}{\sigma_{a1}}\fr{\sigma_{2}}{\sigma_{a}} \left(\fr{x_1 \left(\fr{\sigma_{1}}{\sigma_{a1}}\right) + x_2
\left(\fr{\sigma_{2}}{\sigma_{a2}}\right)}{x_1
\left(\fr{\sigma_{2}}{\sigma_{a2}}\right) + x_2
\left(\fr{\sigma_{1}}{\sigma_{a1}}\right)}\right)}
\en(18)
$$
One can check, using the relation (7), that (18) reduces to the right values
at $x_i=1,\; (i=1,2)$ as well as for $x_1=x_2=1/2.$
For the transverse part $\sigma_{et}$ one obtains
$$
\sigma_{et} (\{\sigma\}, \{x\}) = 
\fr{b' - a\left(\fr{\sigma_{1}}{\sigma_{a1}}
\fr{\sigma_{2}}{\sigma_{a2}}\right)
\left(\fr{x_1 \left(\fr{\sigma_{1}}{\sigma_{a1}}\right) + x_2
\left(\fr{\sigma_{2}}{\sigma_{a2}}\right)}{x_1
\left(\fr{\sigma_{2}}{\sigma_{a2}}\right) + x_2
\left(\fr{\sigma_{1}}{\sigma_{a1}}\right)}\right)}{1+ \left(\fr{\sigma_{1}}{\sigma_{a1}}
\fr{\sigma_{2}}{\sigma_{a2}}\right)
\left(\fr{x_1 \left(\fr{\sigma_{1}}{\sigma_{a1}}\right) + x_2
\left(\fr{\sigma_{2}}{\sigma_{a2}}\right)}{x_1
\left(\fr{\sigma_{2}}{\sigma_{a2}}\right) + x_2
\left(\fr{\sigma_{1}}{\sigma_{a1}}\right)}\right)}.
\en(19)
$$
Again, using now the relation (8), one can check that (19) correctly
reproduces boundary values as well as the exact value at equal
concentrations.

\bs

\und{5. Effective medium approximation in a magnetic field}

\bs
In this section we find out for a completness a "magnetic" transformation for 
the traditional effective medium approximation (EMA) for the effective conductivity. The EMA for $\sigma_e$ in inhomogeneous two-phase self-dual systems without a magnetic field has the form \ci{14}
$$
\sigma_{e} (\{\sigma\}, \{x\}) = (x-\fr{1}{2})\sigma_{-}+
\sqrt{(x-\fr{1}{2})^2 \sigma_{-}^2 + \sigma_1 \sigma_2},
\en(20)
$$
where $\sigma_{-} = (\sigma_1 - \sigma_2).$
Then, the primed effective conductivity will be
$$
\sigma'_{ed} (\{\sigma\}, \{x\}) = c\left((x-\fr{1}{2})\sigma_{a-} +
\sqrt{(x-\fr{1}{2})^2 \sigma_{a-}^2 + \fr{\sigma_1}{\sigma_{a1}} \fr{\sigma_2}{\sigma_{a2}}}\right),
\en(21)
$$
here $\sigma_{a-} = \fr{\sigma_1}{\sigma_{a1}} - \fr{\sigma_2}{\sigma_{a2}}.$
Substituting (21) into (4), one obtains for the diagonal part $\sigma_{ed}$ the next expression
$$
\sigma_{ed} (\{\sigma\}, \{x\}) =
\fr{(a+b')\left((x-\fr{1}{2}) \sigma_{a-} + \sqrt{ (x-\fr{1}{2})^2
\sigma_{a-}^2 + \fr{\sigma_{1}}{\sigma_{a1}} \fr{\sigma_{2}}{\sigma_{a2}}}\right)}{1+ \left((x-\fr{1}{2})\sigma_{a-}
 + \sqrt{ (x-\fr{1}{2})^2
\sigma_{a-}^2 + \fr{\sigma_{1}}{\sigma_{a1}} \fr{\sigma_{2}}{\sigma_{a2}}}\right)^2}.
\en(22)
$$
One can check, using the relation (7), that (22) correctly
reduces to $\sigma_i,$ when $x_i =1$ and to the exact formula for
$\sigma_e$ at $x_1=x_2=1/2.$ For example, at $x_1 = 1, x_2 = 0$ (22)
reduces to
$$
\sigma_{ed} (\{\sigma\},x_1 = 1) = \fr{(a+b') \sigma_1 \sigma_{a1}}{\sigma_{a1}^2 + \sigma_{1}^2}
= \sigma_1.
$$
Analogously, substituting (21) into (4), one obtains for the transverse part of the effective conductivity
$$
\sigma_{et} (\{\sigma\}, \{x\}) = 
\fr{b'- a\left((x-\fr{1}{2})\sigma_{a-} + \sqrt{ (x-\fr{1}{2})^2
\sigma_{a-}^2 + \fr{\sigma_{1}}{\sigma_{a1}} \fr{\sigma_{2}}{\sigma_{a2}}}\right)^2}{1+ \left((x-\fr{1}{2})
\sigma_{a-} + \sqrt{ (x-\fr{1}{2})^2 \sigma_{a-}^2
 + \fr{\sigma_{1}}{\sigma_{a1}}\fr{\sigma_{2}}{\sigma_{a2}}}\right)^2}.
\en(23)
$$
Again, using now the relation (8), one can check that (23) correctly
reproduces boundary values as well as the exact value at equal
concentrations.

It follows from the property of the transformation $T_h$ that $\sigma_e$
belongs to the circumference (9) for all phase concentrations $x \in [0,1]$ \ci{9,10}. The effective conductivity maps the concentration segment $[0,1]$ into
the corresponding arc of this circumference, connecting the points $\sigma_1$ and $\sigma_2$. The different expressions for $\sigma_e$
correspond to the different mappings. The comparison of these mappings is presented in \ci{15}. 
\begin{figure}[t]
\begin{tabular}{cc}
\input epsf \epsfxsize=5.5cm \epsfbox{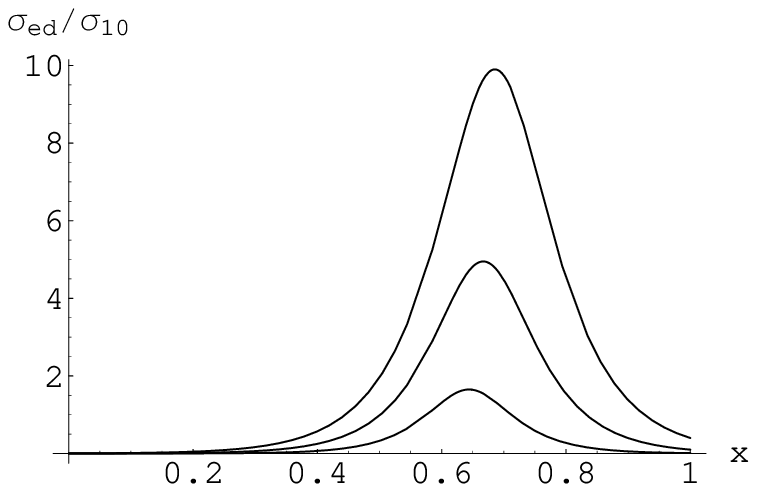}&
\input epsf \epsfxsize=5.5cm \epsfbox{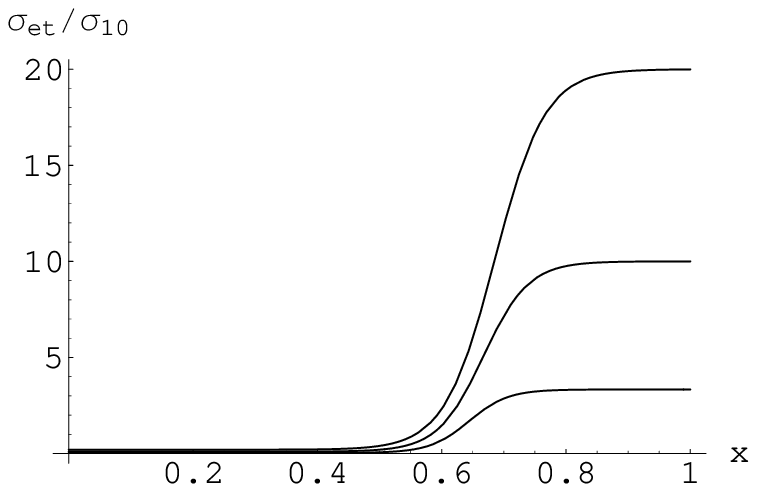}\\
\end{tabular}
\vs{0.3cm}
\cl{a)}
\vs{0.3cm}
\begin{tabular}{cc}
\input epsf \epsfxsize=5.5cm \epsfbox{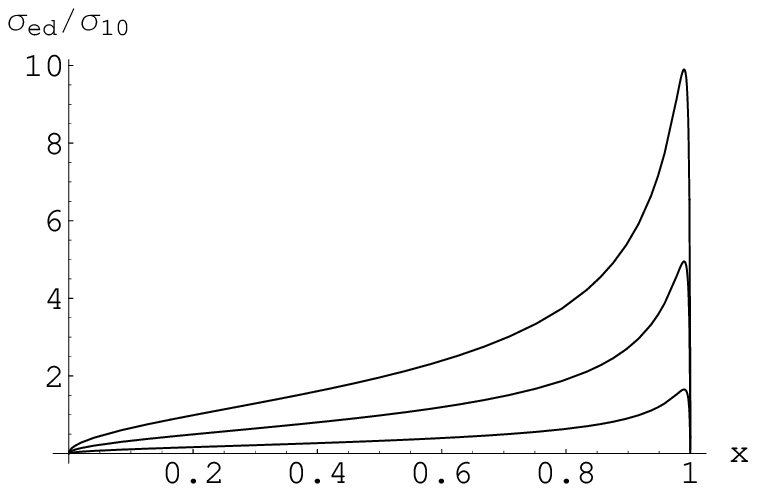}&
\input epsf \epsfxsize=5.5cm \epsfbox{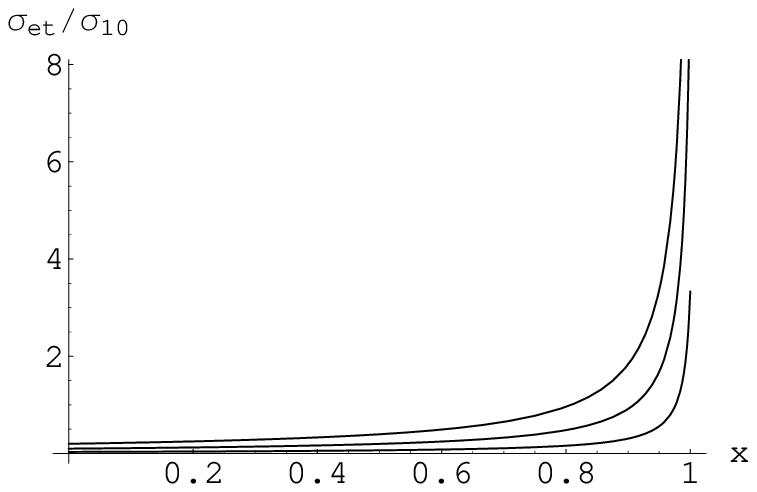}\\
\end{tabular}
\vs{0.3cm}
\cl{b)}
\vs{0.3cm}
\begin{tabular}{cc}
\input epsf \epsfxsize=5.5cm \epsfbox{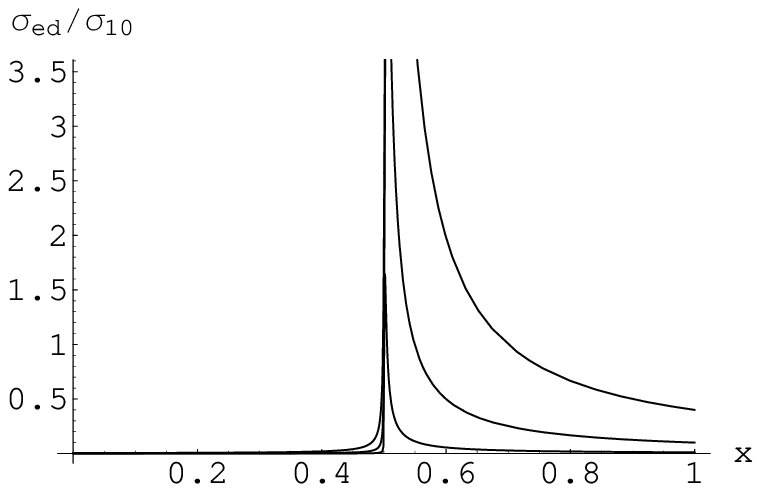}&
\input epsf \epsfxsize=5.5cm \epsfbox{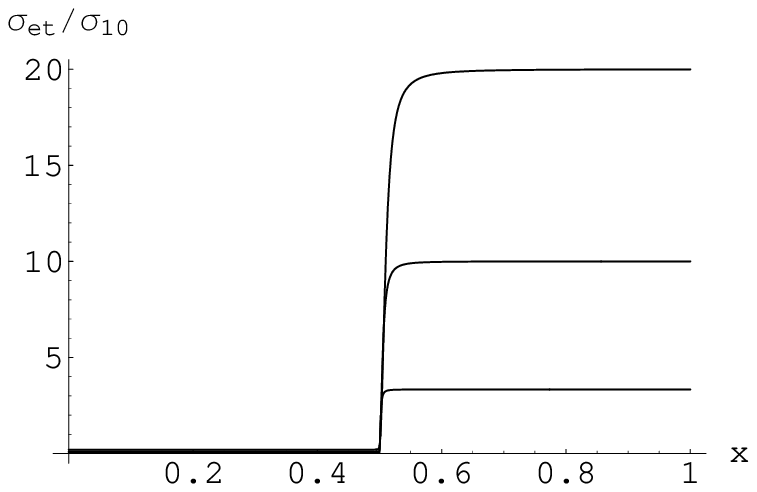}\\
\end{tabular}
\vs{0.3cm}
\cl{c)}
\vs{0.3cm}

{\small  Fig.2.  The  plots of the $x$ dependence of the normalized diagonal 
$\sigma_{ed}/\sigma_{10}$ and transverse $\sigma_{et}/\sigma_{10}$ parts (magnified in 1000 times) for three explicit expressions obtained above (respectively, a),b),c)) at the inhomogeneity
parameter  $\sigma_{20}/\sigma_{10}=0.01,$ and 
at the three different (dimensionless) values of magnetic field $H:$ 1) 50, 2) 100, and 3) 300 (the corresponding plots go from the upper to the lower ones).}
\end{figure}

\bs

\und{6. Effective conductivity in magnetic field}

\bs
Thus, we have found the desired formulas. 
Substituting into (14),(15), (18),(19),(22) and (23) the known functions $\sigma_{id}({\bf H}), \sigma_{it}({\bf
H}), \; (i=1,2),$ one obtains the explicit expressions for the
diagonal $\sigma_{ed}({\bf H})$ and transverse $\sigma_{et}({\bf
H})$ parts of the effective conductivity of inhomogeneous systems in magnetic
field. We give here (see fig.2) their $x$-dependence plots for 2-phase systems, whose partial conductivities in magnetic field can be approximated by the standart (metallic type) formulas \ci{5,3}
$$
\sigma_{id}({\bf H}) = \fr{\sigma_{i0}}{1+ \beta_i^2}, \quad  
\sigma_{it}({\bf H}) = \fr{\sigma_{i0} \beta_i}{1+ \beta_i^2}, \quad 
\beta_i = \mu_i H, \quad i=1,2,
\en(24)
$$
where $\mu_i$ are the corresponding mobilities. 
We will assume here, for simplicity, that $\mu_1 \sim \mu_2 \sim 1.$ 
One can see from fig.2 that the behaviour of the effective conductivities, though different for various expressions, has two common characteristic features:
1) absolute values of $\sigma_{ed}$ are very small, even at high peaks; 2) all values in the regions with noticeable values of $\sigma_e$ decrease with a growth of $H$ at relatively large $H$ approximately linearly. Since $\sigma_{id}$ decrease 
$\sim H^{-2},$ the second feature effectively induces a narrowing of the peaks, which are rather asymmetric and depending on structure of inhomogeneities (i.e. on the form of explicit expressions).
Both these properties are qualitatively compatible with the experimental results
from \ci{1,2}, which show a large magnetoresistivity and its approximately linear growth with an increase of $H.$ More detailed analysis and comparisons with experimental results will be presented in the subsequent papers.

\bs

\und{7. Conclusion}
\bs

Using the exact duality transformation we have found three explicit
approximate expressions for the effective conductivity of 2D isotropic two-phase systems in a magnetic field. The three plots of the dependence of $\sigma_{ed,et}$ on the phase concentration at the inhomogeneity parameter $\sigma_{20}/\sigma_{10}= 0.01$ and
at different values of magnetic field  are constructed. They show very interesting
behaviour of $\sigma_{ed}({\bf H})$ and $\sigma_{et}({\bf H})$, which is qualitatively compatible with the experimental data from \ci{1,2}.
The obtained results can be applied for a description of $\sigma_e$ of 
various two-phase systems (regular and nonregular as well as
random), satisfying the symmetry and self-duality conditions, in a wide range of partial conductivities and at arbitrary concentrations and magnetic fields.
All these results admit a direct experimental checking. 

\bs
\und{ Acknowledgments}

\bs
The authors are thankful to Prof.A.P.Veselov for very useful
discussions of some mathematical questions. This work was
supported by the RFBR grants 00-15-96579, 02-02-16403, and by the
Royal Society (UK) grant 2004/R4-EF.

\end{document}